\documentclass[aps,prl,preprintnumbers,twocolumn,superscriptaddress,groupedaddress]{revtex4}  % for review and submission
\usepackage{graphicx}  % needed for figures
\usepackage{dcolumn}   % needed for some tables
\usepackage{bm}        % for math
\usepackage{amssymb}   % for math
\usepackage{subfigure}
\usepackage{color}
\usepackage{xcolor}
\usepackage{amsmath}
\begin{document}
% The following information is for internal review, please remove them for submission
\widetext
%\leftline{Version xx as of \today}
%\leftline{Primary authors: Hirokazu Maruoka}
%\leftline{To be submitted to PRL}
%\leftline{Comment to {\tt d0-run2eb-nnn@fnal.gov} by xxx, yyy}
%\centerline{\em D\O\ INTERNAL DOCUMENT -- NOT FOR PUBLIC DISTRIBUTION}
% the following line is for submission, including submission to the arXiv!!
%\hspace{5.2in} \mbox{Fermilab-Pub-04/xxx-E}
\preprint{arXiv:1906.05060}
\title{Intermediate asymptotics on dynamical impact of solid sphere on mili-textured surface}
\author{Hirokazu Maruoka}
\affiliation{United Graduate School of Agricultural Science, Tokyo University of Agriculture and Technology, 3-5-8 Saiwai-cho, Fuchu-shi, Tokyo 183-8509, Japan}
\affiliation{Graduate School of Science and Technology, Kwansei Gakuin University, 2-1 Gakuen, Sanda-shi, Hyogo 669-1337, Japan}
\email{maruoka.hirokazu@kwansei.ac.jp}
\email{hmaruoka1987@gmail.com}
%\input author_list.tex       % D0 authors (remove the first 3 lines
                             % of this file prior to submission, they
                             % contain a time stamp for the authorlist)
                             % (includes institutions and visitors)
\date{\today}
\begin{abstract}
Complex phenomena incorporating several physical properties are abundant while they are occasionally revealing the variation of power-law behavior depending on the scale. In this present work, the global scaling-behavior of dynamical impact of solid sphere onto elastic surface is described. Its fundamental dimensionless function was successfully obtained by applying the dimensional analysis combined with the solution by energy conservation complementally. It demonstrates that its power-law behavior is given by the competition between two power-law relations representing inertial and elastic property respectively which is strengthened by scale size of sphere. These factors are successfully summarized by the newly defined dimensionless parameters which gives two intermediate asymptotics in different scale range. These power-law behaviors given by the theoretical model were compared with experimental results, showing good agreement. This study supplies the insights to dimensional analysis and self-similarity in general.
\end{abstract}
%\pacs{77.80.bj,46.55.+d}
\maketitle
%\section{\label{sec:level1}First-level heading}
% sections are not used for PRL papers

\section{Introduction}

In the field of mechanics of continua, including rheology, microfluidics and fluid mechanics, phenomena incorporating several physical properties are frequently observed. Viscoelasticity exhibits both fluidity and solidity while a dimensionless number called Deborah number $De=\tau/T$ \cite{Reiner}, which is defined as the ratio of relaxation time of materials $\tau$ and observation time $T$, qualifies the property. $De \ll 1$ qualifies the material as the fluid while $De \gg 1$ leads to the qualification as solid \cite{Barenblatt2014}. Here note that dimensionless numbers represent the proportion between properties or forces which govern the phenomena (e.g. Reynolds number is the ratio between inertial force and viscous force).  In these two cases, the homogeneous physical property can be assumed in each and the problems generally turn to be simple. However, the intermediate scale range reveals characteristic behavior (e.g. viscoelasticity for $De \sim 1$), in which two physical properties are fundamentally mixed, turns to be complicated problems that are occasionally difficult to be formalized and conquered even though they are quite attractive and important for mechanics of continua. 
 
On the other hand, these phenomena can be understood as {\it intermediate asymptotics} \cite{Barenblatt2006,Barenblatt1972}, which are defined as an asymptotic representation of a function valid in a certain range of independent variables. They are occasionally found as simple power-law relation through dimensional analysis when some dimensionless parameters are considered to be negligible. More or less all the theories can be considered as intermediate asymptotics, which are valid in the certain scale range  \cite{IA}. This concept is formalized by Barenblatt \cite{Barenblatt2014, Barenblatt2006,Barenblatt1972} with the method of dimensional analysis, supplying the universal and coherent view on the physical theory and applications in various area \cite{Banetta,Boscolo,Goldenfeld,Chorin}. This methodology is expected to be effective for the complex problems involving plural physical properties though the scale range in which dimensionless number takes extremely large or small are focused. The method is not always applicable and limited to some extent, particularly in the case where problems turn to be self-similar solutions of second kind. Self-similar solution of second kind is the problem of which dimensionless parameters have power-law behaviors and generally these behaviors cannot be clarified within dimensional analysis but occasionally deduced by some technical manners such as renormalization group theory or method for nonlinear eigenvalue problems. 

The present work focuses on the intermediate scale range of dimensionless parameters in which several physical properties are incorporated, based on the concept of dimensional analysis and intermediate asymptotics. I aim to discuss the relation between dimensionless number and complex behaviors. The problem is dynamical impact of solid sphere onto the mili-textured elastic surface. The dynamical collision is abundant phenomena in our daily life, and interesting for industry \cite{Goryacheva} and sports \cite{Carpick,Nathan}. Since Hertz described the collisional dynamics between two elastic bodies \cite{Hertz}, the theory was developed as contact mechanics \cite{Johnson}. Recently the collision dynamics between macro-textured and immersed sphere is studied by Chastel {\it et al.} \cite{Chastel2016,Chastel2019}. Mili-textured surface can be described by elastic-foundation model \cite{EFM}, of which stress profile is simplified. 

Chastel {\it et al.} have already obtained the scaling behavior of dynamical impact of sphere onto the mili-textured surface. However, I will show that this scaling behavior is an intermediate asymptotic valid in a certain scale range by applying the dimensional analysis. We will recognize the problem belongs to self-similar solution of second kind. Finally I attempt to obtain the fundamental dimensionless functions to describe the global power-law behaviors of this problem by referring to the solution obtained by energy conservation complementally. These theoretical predictions are compared with experimental results to verify the validity of the method.

\section{Experiment}

\begin{figure}[h]
\begin{center}
\includegraphics[width=8.6cm]{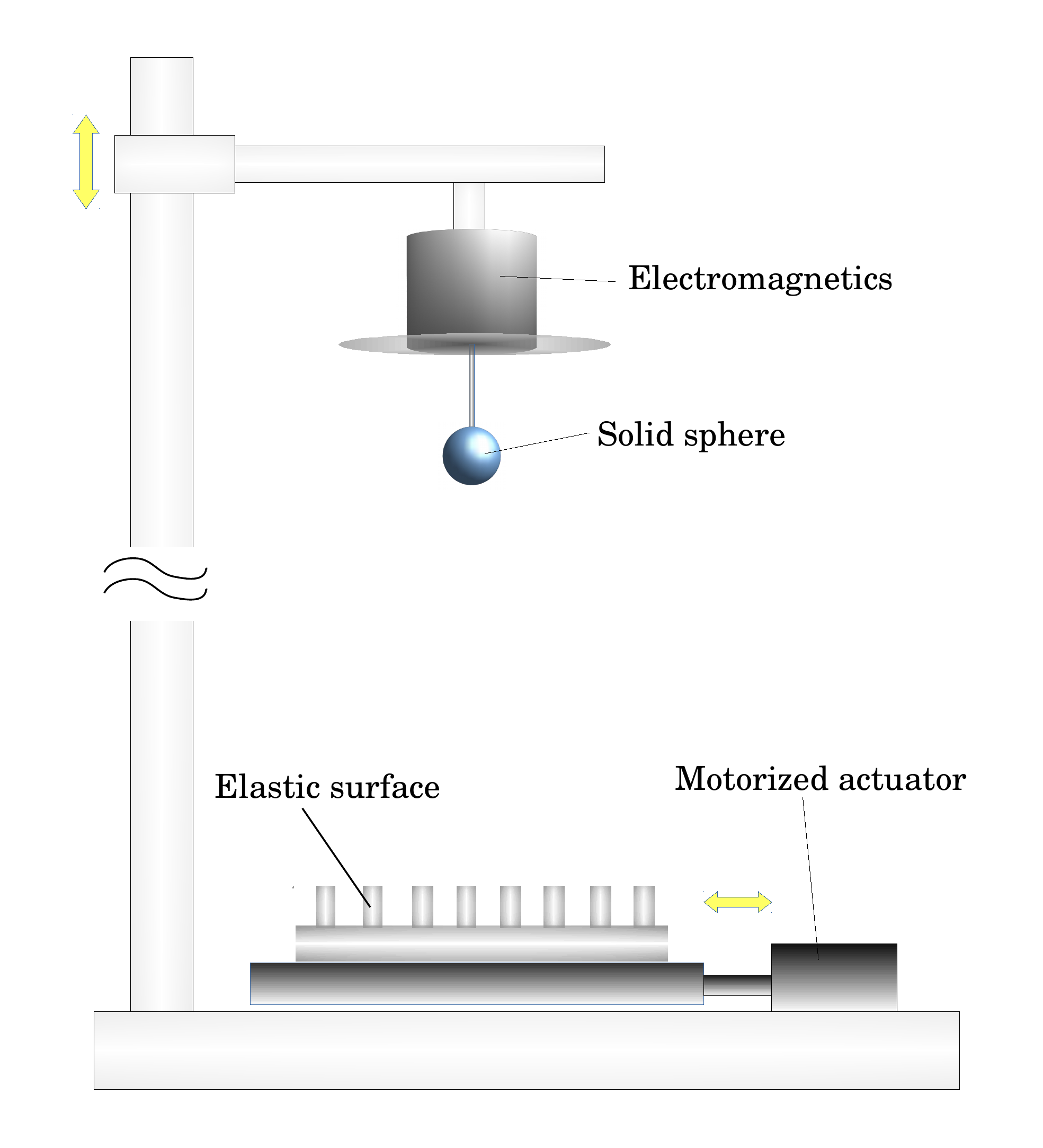}
\caption{(Color online) Sketch of experimental set-up. The solid sphere is suspended by electromagnet which is capable of dropping the ball in arbitrary timing. The velocity of impact can be adjusted by changing the height of the part in which the sphere is suspended. The position of elastic surface made of PDMS can be changed by motorized actuator.}
\label{fig:F1}
\end{center}
\end{figure}

The experiments have been performed using mili-textured surface made of polydimethylsiloxane (PDMS)(kit SYLGARD 180, DOW CORNING) as the elastic surface, of which elastic modulus $E \simeq {\rm 1.6~MPa}$ (see Fig.~\ref{fig:F1}). The periodic, striped-patterned pillars were engraved on the surface, of which the height of pillar ${\rm h = 3.5~mm}$, a square base of side $b={\rm 2.5~mm}$, the interdistance of channel $c={\rm 1.5~mm}$, and the fraction of surface $\phi = b/(b+c) = 0.625$. The metallic sphere (BEARINGOPTION LTD, Steel balls) is suspended by electromagnet (MECALECTRO, F91300 Massy, ${ \rm N^{\circ}}$5,18,01) of which magnet force is controlled and capable of dropping the sphere in arbitrary timing. The collision impacts were recorded by high-speed camera (Phantom V7.3). The collision velocity is varied by changing the height of position from which the sphere is dropped ( $1.5 \sim 50~{\rm cm}$). The size of sphere $R$ is differed as 3.0, 4.0, 4.5, 5.0 and 7.0 mm, of which density $\rho = {\rm 7800~kg \cdot m^{-3}}$. The collision experiments were performed for $30 \sim 40$ times in each conditions, changing the position of the elastic surface every 2.0 mm by motorized actuator so as that the effect of peculiarity between the pillars and sphere were normalized. The information of velocity, deformation and so on was extracted from the movies by image analysis. 

\section{The scaling relation: the result by Chastel et al.}

Firstly, I show the scaling solutions of this problem obtained by Chastel {\it et al.}.

\begin{figure}[h]
\begin{center}
\includegraphics[width=8.6cm]{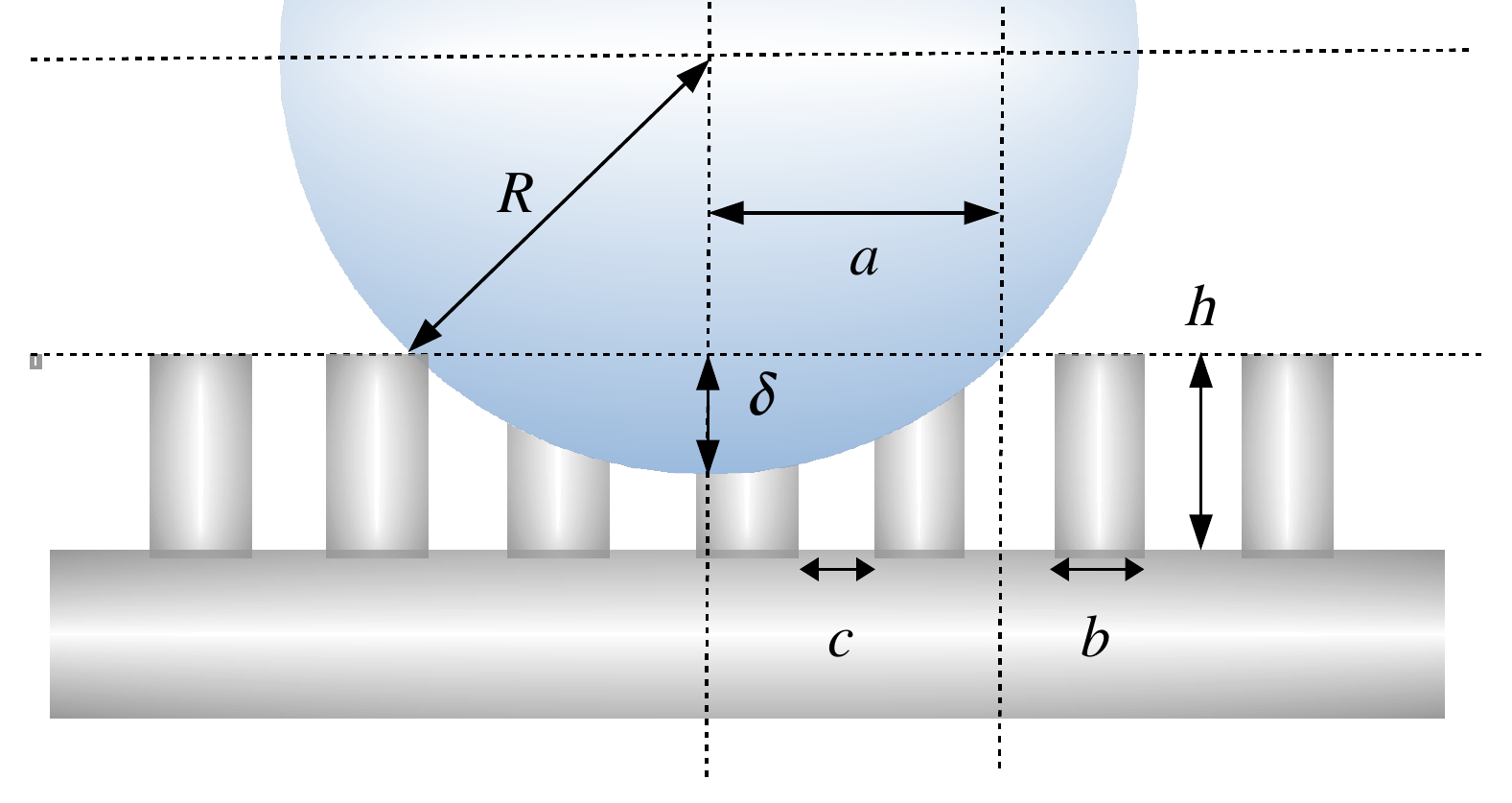}
\caption{(Color online) The geometrical parameters involved in the collision between elastic surface and solid sphere. Deformation $\delta$  and diameter of contact $a$ are generated by the collision onto the elastic PDMS surface.}
\label{fig:F2}
\end{center}
\end{figure}

The collision of sphere falling in the velocity of $v$ onto the surface generating the deformation $\delta$ is sketched on Fig.~\ref{fig:F2}. Assuming Hertzian pressure, $\delta$ is described as $\delta(r)=\delta\left[1-\left(r/a\right)^2\right]$.  According to the theory of Hertz, the contact diameter $a$ is important parameter, which is obtained geometrically, $a^2 = R^2-\left(R-\delta\right)^2 \simeq 2R\delta$. Firstly, the kinetic energy of sphere is easily obtained as 
\begin{equation}
E_{ki}=\frac{2}{3} \pi R^3 \rho v^2.
\label{1a}
\end{equation}
Following the procedure of Chastel {\it et al.} \cite{Chastel2016}, as the normal stress is $\sigma(r)=E\delta(r)/h$, the force of deformation is $F=\int_{0}^{a}\phi \sigma(r) 2 \pi r dr=\pi E \phi R \delta^2 /h$ by eliminating $a$ by $a^2=2R\delta$. Thus elastic energy is obtained as 
\begin{equation}
E_{el}=\int_{0}^{\delta}F(\delta^{'})d \delta^{'}=\frac{\pi E \phi \delta^3 R}{3h}.
\label{1b}
\end{equation}
Thus the conservation equation for kinetic energy and elastic energy at instant $t$ after the collision is described as follows,
\begin{equation}
\frac{2}{3} \pi R^3 \rho v\left(t\right)^2+ \frac{\pi E \phi R \delta\left(t \right)^3 }{3h}= \frac{2}{3} \pi R^3 \rho v^2.
\label{eq:E1c}
\end{equation}
The maximum penetration $\delta$ is reached when $v\left(t\right)=0$, then following relation is obtained,
\begin{equation}
\frac{\delta}{R} = \left(\frac{2}{\phi}\right)^{\frac{1}{3}}\left(\frac{h}{R}\right)^{\frac{1}{3}}\left(\frac{\rho v^2}{E} \right)^{\frac{1}{3}}.
\label{eq:E1}
\end{equation}

Compression time $\tau_c$, which is defined as the duration time at which the sphere contacts with surface \cite{comp}, is obtained as follows,
\begin{equation}
\tau_{c} = 2~\frac{\delta}{v}\int_{0}^{1} \frac{d　(\delta^{'}/ \delta)}{\sqrt{1-(\delta^{'}/\delta)^3 }} = \frac{2}{3}B\left(\frac{1}{3},\frac{1}{2}\right)\frac{\delta}{v}
\label{eq:E1d}
\end{equation}
where $B\left(x,y\right)$ is Beta function. Thus following equation is obtained from Eq.~\ref{eq:E1},
\begin{equation}
\frac{\tau_{c}}{R} =  \frac{C_0}{\phi^{\frac{1}{3}}}\left(\frac{h}{R}\right)^{\frac{1}{3}}\left(\frac{\rho}{E v} \right)^{\frac{1}{3}}
\label{eq:E1e}
\end{equation}
where $C_0 = 2\sqrt[3]{2}/3 \cdot B\left(\frac{1}{3},\frac{1}{2}\right) \simeq 3.533$.

These are the results by Chastel {\it et al}. However, next I show these scaling relations are intermediate asymptotics which are valid in a certain range.

\section{Dimensional analysis : scaling between ${\bf \Pi}$ and ${\bf \eta}$}

Based on the recipe of Barenblatt \cite{FN1}, firstly the function to study $\delta=f(a, R, \rho, E, v, h, \phi)$ is proposed. Assuming $LMT$ unit, the dimensionless parameters are constructed. Here I selected $R,\rho, v$ as the independent parameters, which is the parameters which cannot be represented as a product of the remained parameters. As $\delta = L,~R = L,~a=L, ~\rho = M/L^3, ~E = M/LT^2,~v = L/T,~h=L,~\phi = 1$, following dimensionless parameters are obtained ,
\begin{equation}
\Pi = \frac{\delta}{R},~\xi = \frac{a}{R},~\eta = \frac{\rho v^2}{E},~\kappa = \frac{h}{R}.
\label{eq:E2}
\end{equation}
Thus the function is transformed to $\Pi = \Phi(\xi, \eta, \kappa, \phi)$ where $\Phi$ is an arbitrary dimensionless function. Here let us assume that the problem belongs to the self-similar solutions of second kind \cite{SSSK} as follows,
\begin{equation}
\Pi = \phi^{\gamma_{1}} \kappa^{\gamma_{2}} \eta^{\gamma_{3}} \Phi \left(\xi^{\zeta_1}\kappa^{\zeta_2}\phi^{\zeta_3}\eta^{\zeta_4}\right).
\label{eq:E3}
\end{equation}
Self-similar solutions of second kind are the dimensional analysis solutions which are expressed by the products of dimensionless parameters raised to the powers though the power exponents of dimensionless parameters are not obtained within dimensional analysis in principle. However, in our case, power exponents $\gamma_1\cdots \gamma_3$ can be deduced via Eq.~\ref{eq:E1}. $\zeta_1 \cdots\zeta_4$ are obtained by utilizing Eq.~\ref{eq:E1} and $a^2=2R\delta$, then it is $\xi \sim \phi^{-1/6}\kappa^{1/6}\eta^{1/6}$. Therefore Eq.~\ref{eq:E3} leads to
\begin{equation}
\Pi = \left(\frac{\kappa}{\phi}\right)^{\frac{1}{3}} \eta^{\frac{1}{3}} \Phi \left[\left(\frac{\phi}{\kappa}\right)^{\frac{1}{6}}\xi / \eta^{\frac{1}{6}}\right].
\label{eq:E4}
\end{equation}
Eq.~\ref{eq:E4} is the fundamental dimensionless function which describes the dynamical impact of solid sphere on the mili-textured surface. Supposing new parameters $\Psi=\Pi\phi^{1/3}\kappa^{-1/3}\eta^{-1/3}$ and $\Xi= \xi \phi^{1/6}  \kappa^{-1/6} \eta^{-1/6}$, Eq.~\ref{eq:E4} is described as $\Psi = \Phi(\Xi)$. See $\Psi$ is function with dimensionless parameter $\Xi$. It suggests that the scaling relation derived from the result by Chastel {\it et al.} is confirmed as far as $\Phi$ does not interfere. In this case following intermediate asymptotic is obtained, 
\begin{equation}
\Pi = {\rm const}~\left(\frac{\kappa}{\phi}\right)^{\frac{1}{3}}\eta^{\frac{1}{3}} 
  \label{eq:E5}
\end{equation}
which corresponds to Eq.~\ref{eq:E1}. This condition holds true in the case that $\Xi$ is small enough to consider as $\Phi \sim {\rm const}$. Here we have recognized that $\Xi$ is an important parameter dominating the power-law behavior.  

Next let us move on to the case in which $\Xi$ contributes to the behavior. It is quite interesting to think what kind of intermediate asymptotic is obtained in another scale region. Eq.~\ref{eq:E4} was obtained by the dimensionless parameters via the selection of independent parameters as $R, \rho, v$. However, this choice is arbitrary. Barenblatt suggested that the numerical estimation of dimensionless parameters can help to choose. If the dimensionless parameters to consider is too small or large, these dimensionless parameters can be considered to be negligible. The choice of $R, \rho, v$ is appropriate in the case in which these three parameters play a dominant roll. However the scale range in which $\Xi$ contributes must have $a$ large enough to be considered as $\xi=a/R$ increases $\Xi$. In this case, $a$ should be considered as a dominant parameter.      

Now let us apply the dimensional analysis using another selection of independent parameters as $a, \rho, v$. In this case following dimensionless parameters are finally obtained, 
\begin{equation}
\Pi^{'} = \frac{\delta}{a},~\xi = \frac{a}{R},~\eta = \frac{\rho v^2}{E},~\kappa^{'} = \frac{h}{a}.
\label{eq:E6}
\end{equation}
The difference from Eq.\ref{eq:E2} is that  $\Pi$ and $\kappa$ are replaced by $\Pi^{'}$ and $\kappa^{'}$. Similarly assuming self-similarity of second kind and using Eq.~\ref{eq:E1} and $a^2 \sim R\delta$, the following intermediate asymptotic is obtained in another scale region, 
\begin{equation}
\Pi^{'} = {\rm const} \left(\frac{\xi \kappa^{'}\eta}{\phi}\right)^{\frac{1}{6}}
\label{eq:E7}
\end{equation}
where $\kappa = \xi\kappa^{'}$. Eq.~\ref{eq:E7} is another intermediate asymptotic in the case where $a$ is comparatively large enough. 

Following calculation will justify our interpretation. In order to see the behavior of Eq.~\ref{eq:E4} in wider scale region from small $\Xi$, series expansion of $\Phi$ in power of $\Xi$ is applied as follows,
\begin{eqnarray}
\Pi &=& \left(\frac{\kappa}{\phi}\right)^{\frac{1}{3}} \eta^{\frac{1}{3}} \left\{A_1+A_2~\Xi+A_3~\Xi^2+\cdots\right\} \nonumber \\
  &=& A_1~\left(\frac{\kappa}{\phi}\right)^{\frac{1}{3}} \eta^{\frac{1}{3}} + A_2~\xi\left(\frac{\kappa}{\phi}\right)^{\frac{1}{6}} \eta^{\frac{1}{6}} +A_3~\xi^2+\cdots
\label{eq:E8}
\end{eqnarray}
where $A_1, A_2, A_3$ are constant. Here let us focus on the fact that two dimensionless parameters having different power exponents appear in Eq.~\ref{eq:E8}. Suppose fitting Eq.~\ref{eq:E8} with arbitrary power equation of $\eta$ as follows, $\eta^{\nu} \sim A_1~\phi^{-1/3}\kappa^{1/3}\eta^{1/3} + A_2~\xi\phi^{-1/6}\kappa^{1/6}\eta^{1/6} +A_3 \xi^{2}$, the power exponent $\nu$ is {\it locally} determined and varies in the range $1/6 \leq \nu \leq 1/3$, depending on the contribution of first term and second term in Eq.~\ref{eq:E8}. This balance is critically depends on parameter $\eta$ and $\xi$. We can see that in case of small $\eta$ and large $\xi$, the power exponent of second term 1/6 is dominant. On the other hand, in case of large $\eta$ with small $\xi$, first term is large and dominant, then $\nu$ should be fitted with 1/3. This interpretation corresponds to each intermediate asymptotics Eq.~\ref{eq:E5} and Eq.~\ref{eq:E7} as small $\Xi$ indicates the contribution of second term is extremely small. 

This is spontaneously understood as $\Xi$ is given by ratio of first and second terms as follows, $\xi\phi^{-1/6}\kappa^{1/6}\eta^{1/6} / \phi^{-1/3}\kappa^{1/3}\eta^{1/3} = \xi \phi^{1/6}  \kappa^{-1/6} \eta^{-1/6} = \Xi$. In the end, series expansion of $\Phi(\Xi)$ gives two intermediate asymptotics which are obtained by different selection of independent parameters and $\Xi$ represents the ratio between two intermediate asymptotics.

\section{Dimensional analysis : scaling between ${\bf \tau_c}$ and ${\bf v}$}

Next let us apply the same way to construct the dimensionless function concerning on Eq.~\ref{eq:E1d}. The function to study is $\tau_c=f_{\tau}(a, R, \rho, E, v, h, \phi)$. Assuming the independent parameters as $R,\rho,v$, following dimensionless parameters are to be prepared,
\begin{equation}
\omega = \frac{\tau_c v}{R},~\xi = \frac{a}{R},~\eta = \frac{\rho v^2}{E},~\kappa = \frac{h}{R}
\label{eq:E9}
\end{equation}
to obtain $\omega = \Phi_{\tau}(\xi,\eta,\kappa,\phi)$. Here we assume the self-similar solution of second kind, and we find the following fundamental dimensionless function,
\begin{equation}
\omega = \left(\frac{\kappa}{\phi}\right)^{\frac{1}{3}} \eta^{\frac{1}{3}} \Phi_{\tau} \left[\left(\frac{\phi}{\kappa}\right)^{\frac{1}{6}}\xi / \eta^{\frac{1}{6}}\right]
\label{eq:E10}
\end{equation}
by referring to Eq.~\ref{eq:E1d} and $a^2 = 2\delta R$. Defining $\Omega = \omega\phi^{1/3}\kappa^{-1/3}\eta^{-1/3}$, here we find the relation as $\Omega = \Phi_{\tau}(\Xi)$, suggesting the dependence between $\Omega$ and $\Xi$. Eq.~\ref{eq:E10} gives an intermediate asymptotic corresponding to Eq.~\ref{eq:E1d} as far as $\Xi$ is uninfluential, then we have $\tau_c \sim v^{-1/3}$. However, in the scale range in which $a$ starts to play a roll and $\Xi$ is large enough, another intermediate asymptotic appears,
\begin{equation}
\omega = {\rm const}~\xi\left(\frac{\kappa}{\phi}\right)^{\frac{1}{6}} \eta^{\frac{1}{6}}
\label{eq:E11}
\end{equation}
which is obtained by the series expansion of $\Phi_{\tau}$ as second term, or corresponds to the solution obtained through the dimensional analysis by the selection of the independent parameters as $a,\rho,v$. In this case, scaling relation $\tau_c \sim v^{-2/3}$ appears.
 
\section{Comparison with experimental results}

\begin{figure}[h]
\begin{center}
\subfigure{
 \includegraphics[width=8.0cm]{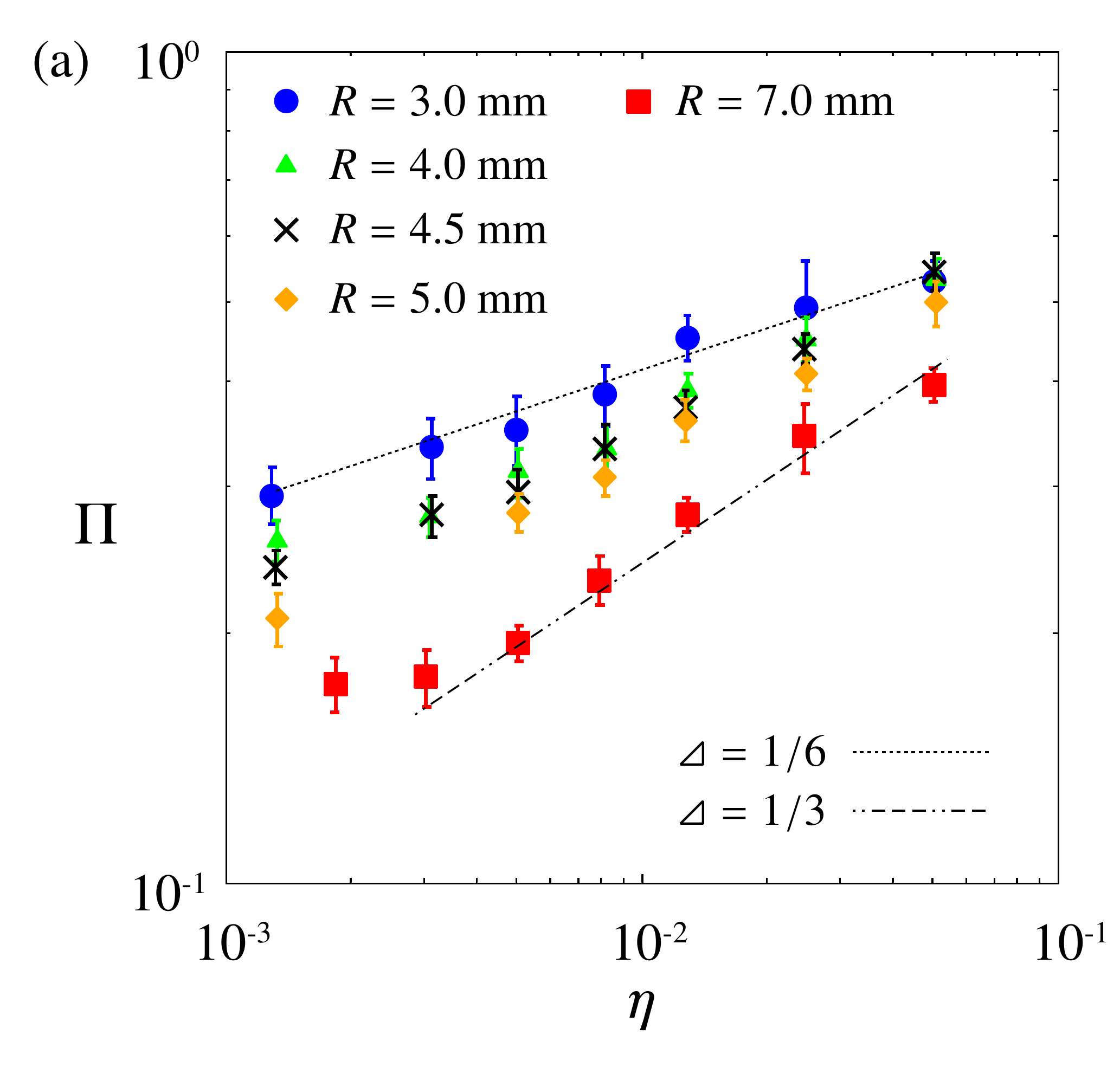}
\label{fig:F3a}}
\subfigure{
 \includegraphics[width=8.0cm]{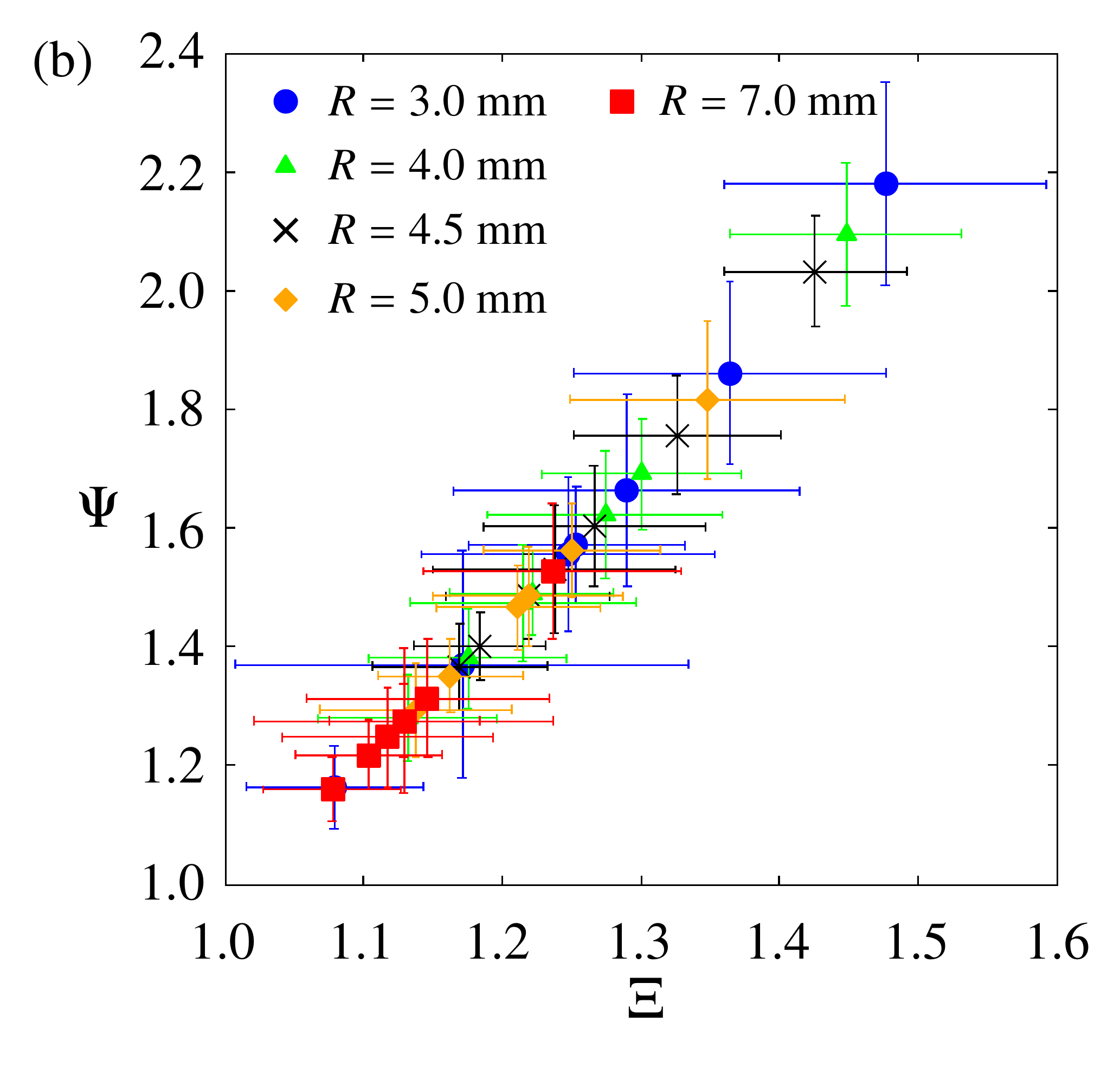}
\label{fig:F3b}}
\caption{(Color online) (a) Power-law relation between $\Pi$ and $\eta$, (b) plots of $\Psi$ vs $\Xi$ in different size of sphere, $R = {\rm 3.0~mm}$ ($\textcolor{blue}{\bullet}$), ${\rm 4.0~mm}$ ($\textcolor{green}{\blacktriangle}$), {\rm 4.5~mm} ($\times$), {\rm 5.0~mm} ($\textcolor{orange}{\blacklozenge}$), {\rm 7.0~mm} ($\textcolor{red}{\blacksquare}$) where $\Pi=\delta/R$, $\eta=\rho v^2/E$, $\Psi=\Pi\phi^{1/3}\kappa^{-1/3}\eta^{-1/3}$ and $\Xi= \xi \phi^{1/6}  \kappa^{-1/6} \eta^{-1/6}$. The two dashed lines indicate the slope of 1/6 and 1/3.}
\end{center}
\end{figure}

Now let us compare these theoretical results with experimental ones. Fig.~\ref{fig:F3a} is the plots of $\Pi$ and $\eta$ in different size of sphere. It is clearly found that the power law behavior varies depending on the size of sphere. Largest sphere $R={\rm 7.0~mm}$ follows the 1/3 power-law behavior, corresponding to Eq.~\ref{eq:E5}. On the other hand, smallest spheres $R={\rm 3.0~mm}$ reveals different power-law behavior, following 1/6 power-law, which corresponds to Eq.~\ref{eq:E7}. 

Fig.~\ref{fig:F3b} is the plots of $\Psi=\Phi(\Xi)$ using experimental data. It is useful to see in which scale range each plots belong to. We can see that plots of small sphere ($R={\rm 3.0~mm}$) which follows 1/6 power-law are belong to larger $\Xi$ while the plots of high velocity decrease $\Xi$. Contrarily it is found that large sphere ($R={\rm 7.0~mm}$) belongs to smaller $\Xi$ though plots of small velocity belong to comparatively larger $\Xi$, which reveals different behavior. Meanwhile, we can find the groups of intermediate size of sphere ($R={\rm 4.0, 4.5, 5.0~mm}$), which belong to $\Xi = 1.1 \sim 1.4$, follow intermediate power-law behaviors. It can be considered as these plots belong to an intermediate scale region in which two power exponents are competing.

\begin{figure}[h]
\begin{center}
\subfigure{
 \includegraphics[width=8.0cm]{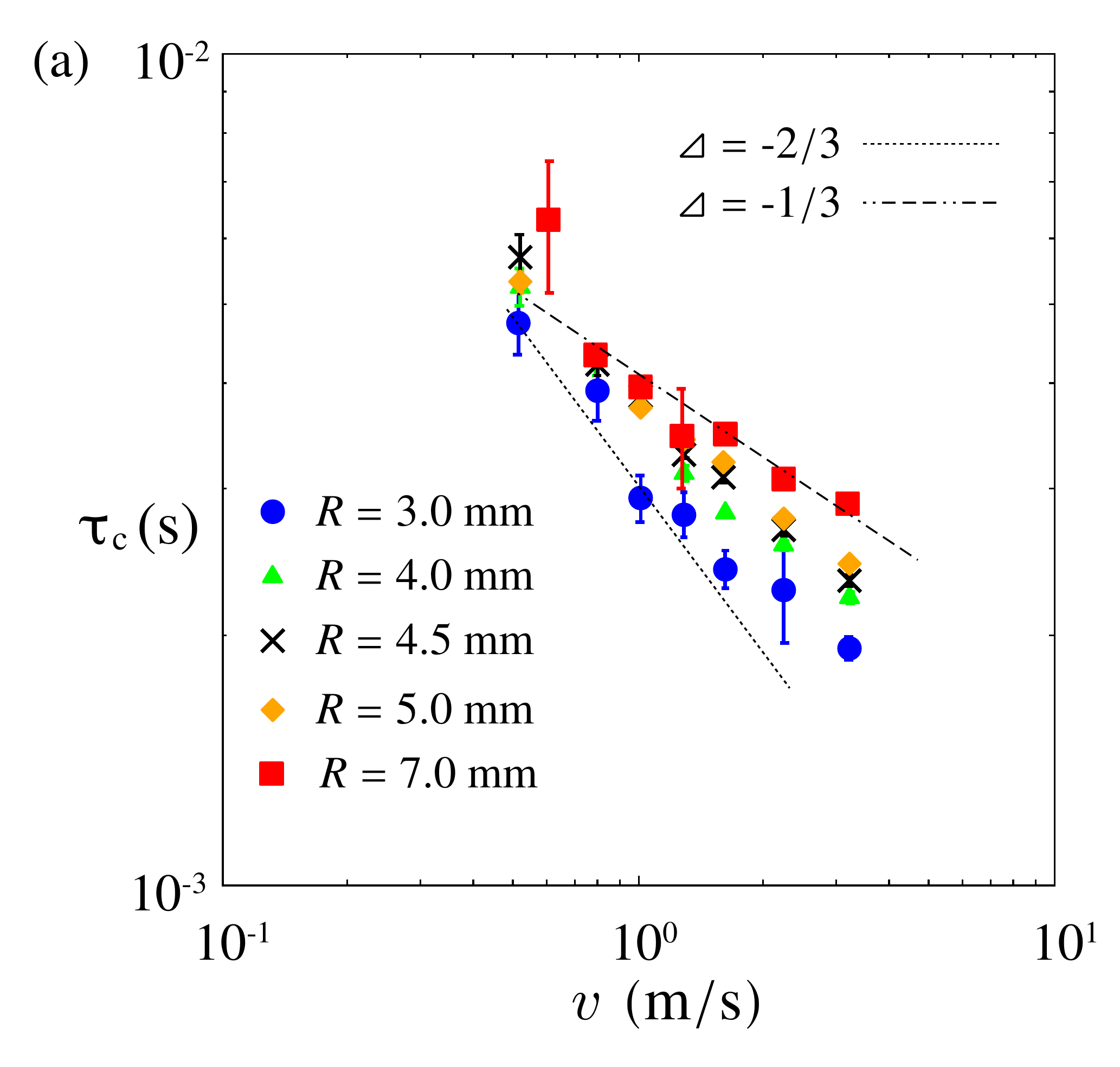}
\label{fig:F4a}}
\subfigure{
 \includegraphics[width=8.0cm]{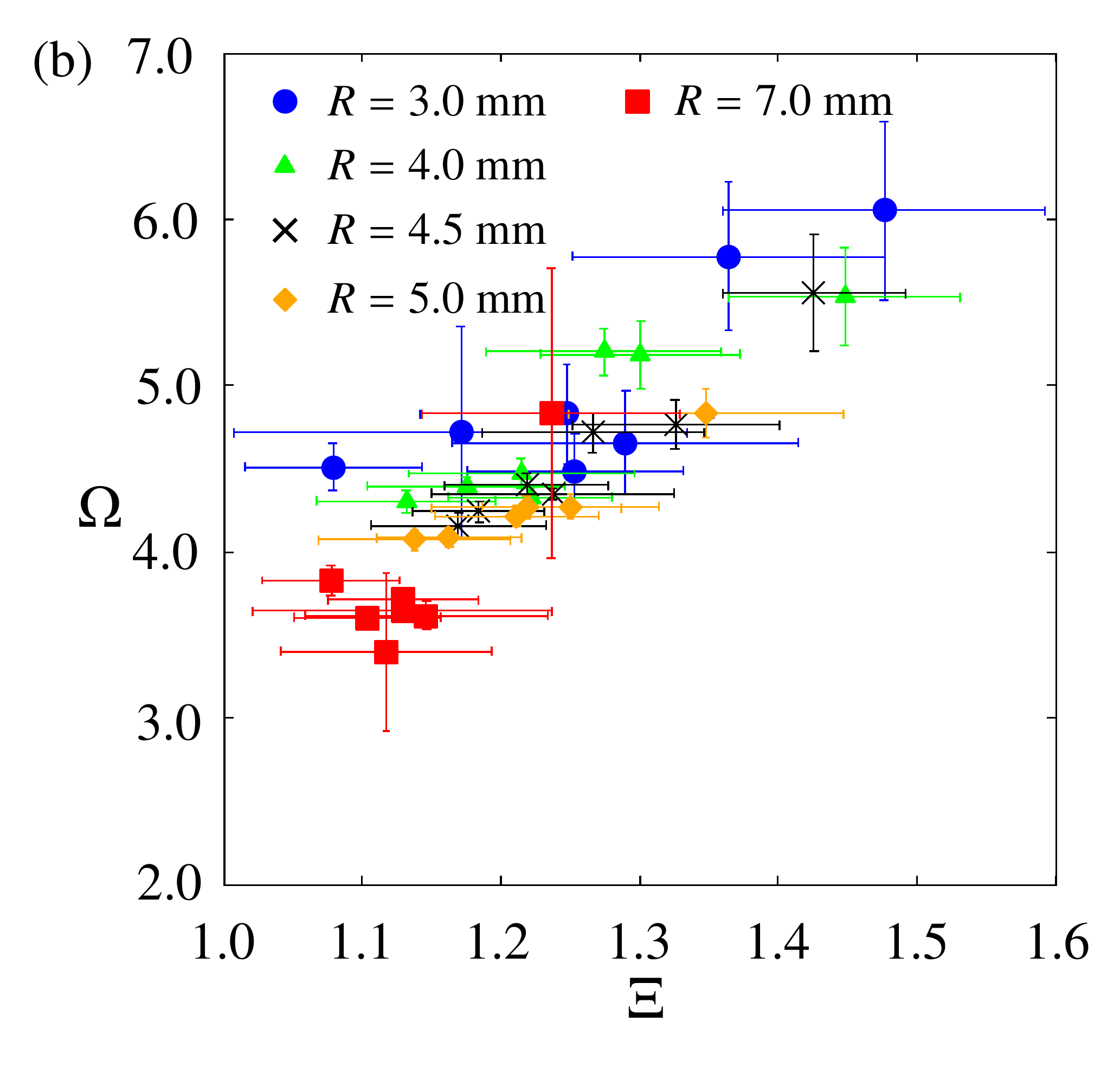}
\label{fig:F4b}}
\caption{(Color online) (a) Power-law relation between $\tau_c$ and $v$, (b) plots of $\Omega$ vs $\Xi$ in different size of sphere, $R = {\rm 3.0~mm}$ ($\textcolor{blue}{\bullet}$), ${\rm 4.0~mm}$ ($\textcolor{green}{\blacktriangle}$), {\rm 4.5~mm} ($\times$), {\rm 5.0~mm} ($\textcolor{orange}{\blacklozenge}$), {\rm 7.0~mm} ($\textcolor{red}{\blacksquare}$) where $\Omega=\omega\phi^{1/3}\kappa^{-1/3}\eta^{-1/3}$ and $\Xi= \xi \phi^{1/6}  \kappa^{-1/6} \eta^{-1/6}$. The two dashed lines indicate the slope of -2/3 and -1/3.}
\end{center}
\end{figure}

The different power-law behavior depending on size of sphere can be seen in the plots of $\tau_c$ and $v$ as well (Fig.~\ref{fig:F4a}). The dimensional analysis predicted two power-law behaviors, $\tau_c \sim v^{-1/3}$ at small $\Xi$ and $\tau_c \sim v^{-2/3}$ at large $\Xi$. The plots of largest sphere ($R={\rm 7~mm}$), having small $\Xi$ as it is shown in Fig.~\ref{fig:F4b}, follows -1/3 power-law behavior which corresponds to Eq.~\ref{eq:E1e}. The plots of smallest sphere ($R={\rm 3~mm}$) reveals mixed behavior though the plots having smaller velocity, belonging to large $\Xi$ in Fig.~\ref{fig:F4b}, follow -2/3 power-law behavior. The plots of the intermediate size sphere follows intermediate behavior.

Focusing on $\Xi$ in detail, we can find that it consists of $\phi$, $\kappa$, $\xi$ and $\eta$. $\phi$ and $\kappa$ are dimensionless parameters which belong to elastic surface, here we focus on the others. $\eta$, which corresponds to Cauchy number that is defined as the ratio of inertial force and elastic force in fluid mechanics, plays a dominant roll on the impact. This parameter reflects the degree of contribution derived from elasticity and inertia. Therefore I would like to call the impact following 1/6 power-law elasticity-dominant impact, and the one following 1/3 power-law inertia-dominant impact \cite{FN2}. 

\begin{figure}[h]
\begin{center}
\includegraphics[width=8.6cm]{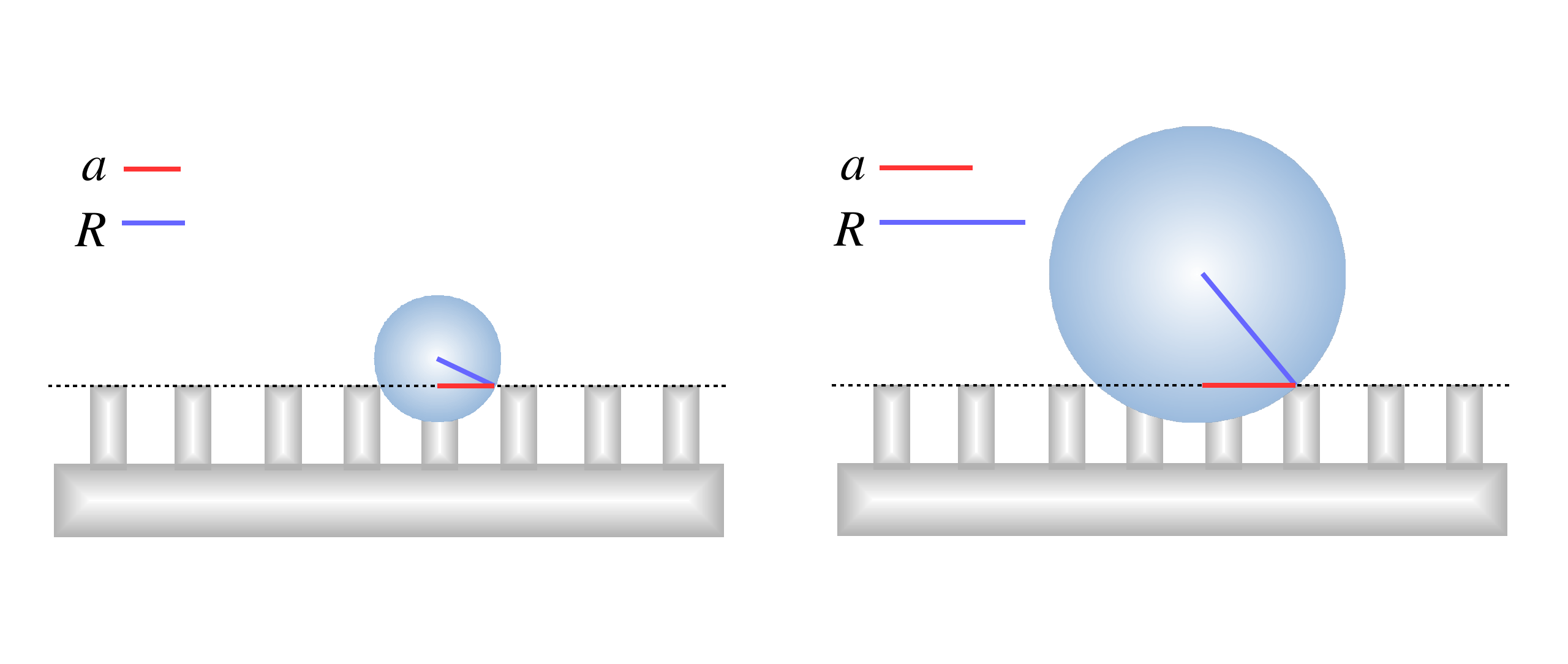}
\caption{(Color online) Comparison of wall/sphere contact generating geometrical dimensionless parameter $\xi=a/R$ in smaller (left) and larger value (right).}
\label{fig:F4}
\end{center}
\end{figure}

Not only $\eta$ but also $\xi$ is a key parameter. The second term of Eq.~\ref{eq:E8}, which corresponds to the intermediate asymptotic of elasticity-dominant impact, is multiplied by $\xi$, indicating that the contribution of second term is critically weakened by small $\xi$. $\xi$ measures the relative degree of subsidence into the surface. Smaller sphere subsides relatively deeper than larger one (Fig.~\ref{fig:F4}), which is the reason why small sphere ($R={\rm 3.0~mm}$) follows 1/6 power-behavior. In the end, these physical interpretation of $\Xi$ corresponds to the analytical interpretation of Eq.~\ref{eq:E8}, which proving the validity of the application of dimensional analysis.

In other reports of contact mechanics, the property of deformation is changed from elastic contact to plastic contact giving different power-laws, depending on the scale of interference \cite{Kogut}. The high speed impact generates the plastic deformation depending on the dimensional parameters \cite{Johnson1972}. The present work discovered another scale-dependent phenomenon of contact mechanics. The scale-dependence of power-law behavior is occasionally observed in self-similarity of second kind \cite{Berry,Barenblatt2002} while the dependence is sometimes semiempirical \cite{Barenblatt1981}. This work clearly identified the dependence of dimensionless parameter as the competition between two power exponents. 

\section{Conclusion} 
 
In conclusion, the above discussion with experimental results confirm the validity of Eq.~\ref{eq:E4} and Eq.~\ref{eq:E10} as the fundamental dimensionless functions of this problem. Eq.~\ref{eq:E4} and Eq.~\ref{eq:E10} include the information of $global$ scaling behaviors, which give two intermediate asymptotics $locally$, depending on $\Xi$. This {\it scale-locality} was quite important to understand this phenomenon as even the power-law behavior depended on the scale. 

The present work is unique on the point that the intermediate scale range in which two physical properties incorporated is focused and the crossover of power-law behaviors are explained as the result of competition between two intermediate asymptotics representing each different physical properties. Generally, the cases in which the uniformity of physical property can be assumed are tend to be concentrated while the intermediate region are avoided. However, this work coped with this intermediate region, and that the crossover of power-law behavior was confirmed with experimental results. Furthermore, the two different method were combined complementally in this work: dimensional analysis and the solution obtained by the equation of kinetic energy and elastic energy. Generally latter solution is considered to be enough but the scale dependence would not have been recognized without dimensional analysis. This suggests that this combinated dimensional analysis with the concept of intermediate asymptotic is quite effective to analyze the mesoscale phenomena incorporating two or more physical properties, revealing different behaviors depending on the scale. 

In this work, self-similarity of second kind is understood as the competition between two intermediate asymptotics. This is also quite interesting insight for the concept of self-similarity in general.\\

\section{Acknowledgement}

The author wishes to thank J.-B. Besnard, P. Panizza and L. Courbin for technical assistance of the experiments. He thanks to K. Osaki for theoretical advice. This work was supported by the program Long-term Internship Dispatch for Innovation Leader Training organized by Building of Consortium for the Development of Human Resources in Science and Technology.


\begin{thebibliography}{99}
\bibitem{FN2}
Simply calling elastic impact or inertial impact may confuse with the term {\it elastic impact} which has already been defined as the collision of which the coefficient of restitution equals 1.
\end{thebibliography}
\end{document}